# Charge confinement by Casimir forces

H. E. Puthoff
Institute for Advanced Studies at Austin
4030 W. Braker Ln., Suite 300
Austin, TX 78759

and

M. A. Piestrup
Adelphi Technology, Inc.
981-B Industrial Rd.
San Carlos, CA 94070

**Abstract**

Laboratory observation of high-density filamentation or clustering of electronic charge suggests that under certain conditions strong coulomb repulsion can be overcome by cohesive forces as yet imprecisely defined. Following an earlier suggestion by Casimir, we investigate here the possibility that Casimir forces can lead to charge clustering of the type observed, and conclude that such forces may play a role in the generation of robust high-charge-density effects.

## I.   Introduction

The formation of high-density, electronic charge clusters has been reported to occur under certain precisely-defined laboratory conditions, ranging from quantum dots for nanotechnology applications [1], through high-power-density microelectronic devices utilizing micro-arc discharge processes [2,3], to the containment of plasma for fusion purposes [4].

Aside from the trivial case of electron charge neutralization by positive ions, mechanisms proposed in the literature for high-density charge confinement range from standard magnetic pinch models to exotic, soliton-like localized-wave (LW) solutions in plasma-EM wave interactions [5].

Another candidate mechanism that has yet to be fully explored with regard to high-density charge confinement is provided by the short-range, attractive van der Waals and Casimir forces driven by vacuum-fluctuation phenomena. Such forces derive from the fact that the vacuum, rather than being the void of classical theory, is the seat of electromagnetic zero-point fluctuations (ZPF) comprising significant energy density and radiation pressure. Theory predicts and experiments verify the existence of ZPF-driven van der Waals-type forces between closely-spaced metallic or dielectric boundaries, and between free-charge distributions. One particular example is the Casimir force, the attractive ZPF-induced quantum force between conducting parallel plates [6-9], recently measured with good precision [10,11]. Detailed analysis shows that the attractive force can be modeled as due to partial shielding of the ZPF radiation pressure from the interior region of the plates which results in the plates being pushed together [12].

As to the issue of charge confinement specifically, it was Casimir himself who first proposed that the ZPF-induced pressure forces might provide a mechanism for the confinement of charge in (semi-classical models of) elementary particles [13], a problem even more challenging than the laboratory observations under consideration here. To explore this concept we examine the consequences of collective behavior of multiparticle charge distribution in the geometries proposed by Casimir.

## II.     Multiparticle van der Waals/Casimir-Force Interaction

Early investigation by Casimir of vacuum-fluctuation-driven, multiparticle van der Waals interactions led to the realization that summation over the collective particle-particle interactions could be recast to advantage in terms of the restructuring of vacuum energy in response to the boundaries defined by the particle distribution [6]. This is because charge distributions which can respond to fields constitute boundaries which reconfigure those fields, vacuum ZPF fields included. The classic example is the Casimir Effect itself in which parallel conducting plates exclude all but a finite number of modes in the interior spacing, with the consequence that the radiation pressure outward on the plates associated with the interior modes is less that the pressure inward due to the (essentially infinite in number) external modes. This yields a net radiation pressure that drives the plates together with a force per unit area given by [12]

$$\frac{F}{A} = -\frac{\pi^2}{240}\frac{\hbar c}{d^4},$$

where $d$ is the spacing between the plates.

With regard to the problem of interest here, the containment of high-density charge, Casimir suggested that a shell-like distribution of charge might partially shield vacuum fields from the interior of the shell with the result that net inward pressure would compensate outwardly-directed coulomb forces to yield a stable configuration at small dimensions. Along these lines Casimir offered two models for consideration [13], one in which interior fields associated with discrete states permitted by boundary conditions are assumed to exist as in the parallel-plates example above (Casimir Shell Model I), and a second which assumes total shielding of interior fields up to a Compton-frequency cutoff for electron-ZPF field interactions (Casimir Shell Model II).

## III. Casimir Shell Models

In Casimir's Shell Model I, interior fields associated with discrete states permitted by boundary conditions are assumed to exist. One of the first applications of this model to charge confinement considered a semiclassical electron as a conducting spherical shell carrying a homogeneously-distributed surface charge $e$ whose tendency to expand by coulomb repulsion is checked by inwardly-directed ZPF radiation pressure (the Casimir force). Unfortunately for the model, a detailed analysis found that for this case the Casimir pressure was outwardly-directed, augmenting rather than canceling the coulomb pressure [14-17].

With regard to Casimir's Shell Model II (total shielding of interior fields up to some cutoff frequency) its plausibility finds some degree of support in work by one of the authors (H.P.) that traces the source of the ZPF fields to the quantum-fluctuation motion of charged particles distributed over cosmological space [18]. Such fields could then in principle be shielded from an interior space by a sufficiently dense charge distribution. It is this second model that shows some promise.

With the spectral energy density of the ZPF fields given by

$$\rho(\omega) = \frac{\hbar \omega^3}{2\pi^2 c^3} d\omega,$$

integration over frequency from zero up to a cutoff frequency $\Omega_c$ for electron-ZPF interactions yields a vacuum energy density $u_V$ effective in such interactions

$$u_V = \int_0^{\Omega_c} \rho(\omega) d\omega = \frac{\hbar \Omega_c^4}{8\pi^2 c^3}.$$

Now assume a spherical-shell distribution of $N$ electrons on a shell of radius $a$. For the ZPF radiation pressure to compensate the electrons' coulomb stress we require $(1/3) u_V = u_e$, or

$$\frac{1}{3} \frac{\hbar \Omega_c^4}{8\pi^2 c^3} = \frac{1}{2} \varepsilon_o E^2 = \frac{N^2 \alpha \hbar c}{8\pi a^4},$$

where $\alpha$ is the fine-structure constant, $\alpha = e^2/4\pi\varepsilon_o \hbar c \approx 1/137.036$. This leads to an expression for the cutoff frequency[1]

---

[1] Alternatively, one can derive the expression by equating the ZPF energy excluded from the interior of the sphere during adiabatic formation to that now stored in the coulomb field.

$$\Omega_c = (3\pi\alpha)^{1/4} \sqrt{N}\left(\frac{c}{a}\right).$$

With regard to the cutoff frequency $\Omega_c$ for electron-ZPF interactions, we choose the Compton frequency $\hbar\Omega_c = m_e c^2$, where $m_e$ is the electron mass. This choice for the cutoff has been shown to yield correct results for other electron-ZPF interactions, for example in calculations of the Lamb shift [19,20]. Substitution into the above equation then yields an expression for the diameter $d$ of the spherical-shell distribution,

$$d = 2a = 2(3\pi\alpha)^{1/4} \sqrt{N}\left(\frac{\hbar}{m_e c}\right) \approx \sqrt{N}\left(\frac{\hbar}{m_e c}\right).$$

We see that the diameter $d$ for the spherical-shell distribution predicted by this calculation is simply the Compton wavelength $\hbar/m_e c$ multiplied by the square root of the number of electrons on the shell. Such numbers are in rough correspondence with claimed observations of charge clustering phenomena.

As an additional check on this simple model we examine whether the so-called *Schrödinger pressure* plays a confounding role in this configuration.[2] The expression for the Schrödinger pressure exerted by $N$ electrons confined to a volume $V$ in their lowest state ($N/2$ with spin up, $N/2$ with spin down) is [21]

$$P_s = \frac{1}{5}(3\pi^2)^{2/3} \frac{\hbar^2}{m_e}\left(\frac{N}{V}\right)^{5/3}.$$

A direct comparison of the magnitude of the Schrödinger pressure against the vacuum and coulomb pressures given by $(1/3)u_V$ and $u_e$, respectively, indicates that the Schrödinger pressure can be neglected for electron numbers of interest here, say $N > 10^4$.

Finally, we note that the simple spherical-shell configuration analyzed here satisfies a first-order stability condition that, once configured, a slight change in radius results in a restoring force that tends to return the sphere to its original size.

IV.  **Conclusions**

Laboratory observation of high-density filamentation or clustering of electronic charge has motivated an investigation into potential cohesive mechanisms whereby repulsive coulomb forces could be overcome by some form of compensatory force of attraction. Of the various possibilities discussed in the literature, we have chosen to examine a model suggested by Casimir that invokes the possibility of charge confinement by Casimir forces. The resulting analysis indicates that the cooperative action of large numbers of charges by Casimir-type effects does provide a potential candidate for charge confinement of roughly the right order of magnitude to correlate with reported laboratory observations.

---

[2] The Schrödinger pressure is a force that resists particle confinement due to the wave nature of matter that derives from a combination of the uncertainty and Pauli Exclusion principles.

## Acknowledgements

This work was sponsored in part by the Air Force Office of Scientific Research, Bolling AFB, DC.